\documentclass[twocolumn,pre,showpacs]{revtex4}              
\usepackage{hyperref}                                        
\usepackage{graphicx}        
\usepackage{amssymb,amsfonts,amsmath}                        
\usepackage{bm}                                              
\usepackage{dcolumn}                                         
\usepackage{color,colordvi}                                  

\usepackage{soul}					     

\usepackage{psfrag}					     

\begin{document}


\title{Modular Dynamical Semigroups for Quantum Dissipative Systems}

\author{David Taj}
\email{david.taj@mat.ethz.ch}
\author{Hans Christian \"Ottinger}
\affiliation{Polymer Physics, Department of Materials, Eidgen\"ossische Technische Hochschule Z\"urich, CH-8093 Z\"urich, Switzerland}%

\begin{abstract}
We introduce a class of Markovian quantum master equations, able to describe the dissipative dynamics of a quantum system weakly coupled to one or several heat baths. The dissipative structure is driven by an entropic operator, the so called modular Hamiltonian, which makes it nonlinear. The generated Modular Dynamical Semigroup (MDS) is not, in general, a Quantum Dynamical Semigroup (QDS), whose dynamics is of the popular Lindblad type. The MDS has a robust thermodynamic structure, which guarantees for the positivity of the time evolved state, the correct steady state properties, the positivity of the entropy production, a positive Onsager matrix and Onsager symmetry relations (arising from Green-Kubo formulas). We show that the celebrated Davies generator, obtained through the Born and the secular approximations, generates a MDS. By unravelling the modular structure of the former, we provide a different and genuinely nonlinear MDS, not of QDS type, which is free from the severe spectral restrictions of the Davies generator, while still being supported by a weak coupling limit argument. With respect to the latter, the present work is a substantial extension of~\cite{Ottinger2011_GEO,Ottinger2010_TLS_DHO}.
\end{abstract}





\maketitle

Quantum Markovian master equations~\cite{Davies1974,Gorini1978149,Alicki2007,Breuer2002} are being used since half a century to model the dynamics of a quantum system in contact with a particle reservoir~\cite{Davies1974,Davies1976,Davies1975}. If on the one side they are justified on the statistical mechanical grounds of an underlying purely Hamiltonian system in a scaling limit~\cite{Davies1974,Davies1976}, their interest often stems from their own structural properties, particularly with respect to thermodynamic requirements, on the other~\cite{SpohnLeb,Spohn1978,Alicki_1977eu}. A famous example is the Davies master equation~\cite{Davies1974}, obtained through the Born plus secular approximations~\cite{Breuer2002} and giving rise to a Quantum Dynamical Semigroup through a generator of the popular Lindblad kind~\cite{Lindblad1976}. As many other known generators, the Davies generator can (i) describe the Hamiltonian evolution of a small subsystem interacting with a particle reservoir at equilibrium in the weak-coupling limit~\cite{Davies1976,Taj2010}. The Davies generator turned out to be the most successful and popular choice, as it comes with the additional advantages of guaranteeing (ii) the complete positivity of the time evolved state~\cite{gorini76}, (iii) the correct steady state~\cite{Davies1974}, (iv) the positivity of the entropy production, (v) a positive Onsager matrix~\cite{DegrMaz} and (vi) Onsager symmetry relations for heat baths in the linear regime (arising from Green-Kubo formulas)~\cite{Kubo1991,SpohnLeb,Jakia_2006kl}. That all these properties are fulfilled for finite values of the coupling to the bath, and not just asymptotically in the scaling limit, has given the status of a physically robust quantum dissipative dynamics per se to the Davies generator.

However, Davies himself in his foundation papers~\cite{Davies1974,Davies1976} pointed out that his generator suffered from severe spectral limitations, and could not even be defined for unbounded subsystems. In~\cite{Taj2008,Taj2010} this problem was studied, and a cut-off dependent Lindblad generator was proposed. The generator was supported by the same weak-coupling argument as for the Davies case, and unlike the latter it did not suffer from any spectral restriction. However, nothing could be said in general for properties (iii)-(vi) above, at finite coupling to the bath. Therefore, a master equation able to extend the Davies generator beyond its spectral limitations, while preserving (i)-(vi) above, is still lacking. Moreover, even for bounded subsystems, the important question remains, as to weather the Davies master equation is unique: Is there really no other possible Markovian dynamics left, which guarantees (i)-(vi) above?

In this paper, we shall present a class of nonlinear Markovian master equations, describing the dissipative time evolution of out of equilibrium excitations of a quantum system coupled to one or several heat baths. The scattering event governing the irreversible dynamics is driven by an entropic operator at the origin of the nonlinearity, known as the modular hamiltonian. For this reason we say that our dynamics generates a Modular Dynamical Semigroup~(MDS). We prove a thermodynamic analogue of complete positivity (the latter notion is not even available for a nonlinear dynamics), namely that (ii') the relative entropy operator is bounded at all positive times, even when an ancilla system is included. Properties (ii'-vi) are all consequences of the modular structure, as we show. MDSs are at variance with the linear QDSs. We find that the only QDS which is also a MDS is that given by the Davies generator. Unravelling the modular structure of the latter immediately leads to a possible nonlinear MDS extension, able to describe unbounded systems and systems in the thermodynamic limit. That the generalization is not unique gives substantial freedom for engineering novel dissipative quantum phenomena through a simple, Markovian master equation. Being now able to properly describe systems in the weak coupling regime, this work is a structural extension of~\cite{Ottinger2011_GEO,Ottinger2010_TLS_DHO}. As therein, more general environments can be taken into account, although here we only consider heat baths for sake of exposition.

We take $ \hbar =1$ and let the states of the quantum system be described by strictly positive and normalized trace class operators  $\rho$ over a Hilbert space $\mathcal H$. Observables $A$ are bounded operators on $\mathcal H$. The system is put in contact with a heat bath at inverse temperature~$\beta$ and otherwise undergoes a free hamiltonian dynamics generated by a Hamiltonian $H_S$ (potentially containing renormalization terms). The irreversible influence of the heat bath on the system is implemented through a bounded scattering operator $Q^\lambda$ depending on $\lambda\in[0,1]$ and on $\beta$. 

We let $\Delta S =- \ln \rho - \beta H_S$ be the relative entropy operator. 
We define a $(\beta,\rho)$-dependent modular dissipative bracket $[[\cdot,\cdot]]_{\beta,\rho}$, as the positive bilinear form over the observables
\begin{equation}\label{disbra}
[[A,B]]_{\beta,\rho} \!=\!\! \int_0^1\!\! \left(\zeta^{i\lambda\beta \over 2}_\rho (i [Q^\lambda, A^\dagger])\right)^\dagger  \!\! \zeta^{i\lambda\beta\over 2}_\rho(i [Q^\lambda, B])\, d\lambda,
\end{equation}
constructed with the help of the hamiltonian evolution
\begin{equation}\label{modev}
\zeta^t_\rho(A) = e^{i t  (-\beta^{-1}\ln \rho)} A e^{-i t  (-\beta^{-1}\ln \rho)},
 \end{equation} 
generated by the modular hamiltonian $-\beta^{-1}\ln\rho$ at the imaginary time $t = i\lambda\beta/2$ (see e.g.~\cite{bratteli,Pillet2010}).

Through such brackets, and denoting $\rho(A) = {\rm tr}(\rho A)$ when there is no room to confusion, we define the MDS $W_{t,t'}$ to be the evolution map $W_{t,t'}(\rho_{t'}) = \rho_t$ associated to the master equation
\begin{equation}\label{MDS}
\partial_t\rho(\cdot) = \rho\left( i[H_S,\cdot]\right) +\rho\left([[\cdot,\Delta S]]_{\beta,\rho} \right).
\end{equation}
 We note that this is a generalization of the nonlinear quantum master equation presented in~\cite{Ottinger2011_GEO,Ottinger2010_TLS_DHO}, which is recovered in the particular case of $Q^\lambda=Q$ not depending upon $\lambda$ (with the understanding that extension of the present formalism to more general environments than heat baths can be made by letting the environment brackets in~\cite{Ottinger2011_GEO} depend on $\lambda$). This is an important conceptual development, motivated by a much wider applicability regime of the resulting theory, as we shall see. It allows the bath to have a more general kind of influence the system, thanks to what could be named colored modular noise. Indeed, the evolution (\ref{modev}) at the microscopic imaginary time $i\lambda\beta/2$, during the scattering event, is weighted by the coupling operator $Q^\lambda$. On the top of this colored modular noise one could accommodate for colored noise in the usual sense by extending (\ref{MDS}) to many coupling operators $Q^\lambda_\alpha$ for $\alpha\in \mathcal I$ in some energy level set (as we shall later do). The interpretation of (\ref{MDS}) is clearly that the relative entropy operator $\Delta S$ acts as a thermodynamic force driving the system to equilibrium, just as in~\cite{Ottinger2011_GEO} . 

For the well posedness of (\ref{MDS}) it is sufficient to assume, as we do, that $Q^{\lambda \dagger} = Q^{1-\lambda}$ (which guarantees self-adjointness), and that $[Q^\lambda,H_S]$ is bounded. We further make the ergodicity assumption
that $H_S$ and $Q^\lambda$ do not share common eigenspaces~\cite{SpohnLeb,Jaksic2013_QDS}. This guarantees that all scattering channels are coupled, leading to a unique steady state, when the latter exists (in which case it is the equilibrium state $\rho_\beta$, as seen from $\Delta S = c1$).

Note that the modular evolution $\zeta_\rho^t$ in ({\ref{modev}) is still well defined in the thermodynamical limit~\cite{bratteli}, where the positive normalized functional $\rho(\cdot)$ implementing the state of the system can no longer be represented by a density matrix~\cite{Pillet2010}. As a direct consequence the MDS master equation (\ref{MDS}) easily extends to the thermodynamical limit, at least as long as $\Delta S$ is bounded throughout the evolution. Note that if $\mathcal H$ is finite dimensional, as we shall from now on assume to avoid technicalities, this is the case for just any strictly positive $\rho>0$.

This leads us to the following positivity argument. First, (\ref{MDS}) preserves normalization through $\partial_t\rho(1) =0$. The only way for $\Delta S$ to diverge is then for some eigenvalue $p$ of $\rho$ to go to $0$. When this happens, the entropy $S(\rho) = \rho(-\ln\rho)$ decreases. But from
$
\partial_t S(\rho) = \rho([[-\ln\rho,\Delta S]]_{\beta,\rho})
$
and from $[Q^\lambda,\Delta S] = [Q^\lambda,\ln\rho] +O(1)$ (due to the supposed boundedness of $[Q^\lambda,H_S]$), we find that in this limit, the Von Neumann entropy rate is asymptotic to 
\begin{equation}\label{entprod}
\sigma(\rho) = \rho\left(  [[\Delta S,\Delta S]]_{\beta,\rho} \right) \geq 0,
\end{equation}
which is positive (because of the positivity of~(\ref{disbra}), thus guaranteeing that $\Delta S$ stays bounded throughout the evolution. 

We wish at this point to discuss the positivity property of the MDS in comparison to the notion of complete positivity~\cite{gorini76, Lindblad1976} that cannot even be defined for our nonlinear dynamics (which is not even of mean-field type~\cite{alicki1990}). We take the physical essence of the disputed notion (see e.g.~\cite{Shaji200548} and references therein) to be that the presence of a external ancilla quantum system $S'$, non-interacting with $S$, cannot spoil the positivity of the evolution. From our perspective, the most natural implementation of such concept is that the total hamiltonian and the total coupling operator of $S+S'$ should not couple degrees of freedom of $S$ and $S'$. We therefore take them of the form $H_S\otimes 1_{S'} + 1_S\otimes H_{S'}$ and $Q^\lambda_S\otimes 1_{S'} + 1_S\otimes Q^\lambda_{S'}$ respectively. Then the MDS of $S$ naturally extends to an MDS of $S+S'$, which is again positive (and so it stays in the limit of vanishing $H_{S'}$ and $Q^\lambda_{S'}$, provided that ergodicity is guaranteed throughout the limit). This is why we can safely state that MDS are completely positive in this sense. We believe that our alternative and inequivalent notion of complete positivity, although trivial and innocent, can shine light on the problematic conventional definition, which itself exhibits such surprisingly drastic physical consequences on the system $S$~\cite{Shaji200548}.

Actually there is a deeper thermodynamical meaning to (\ref{entprod}): by defining the mean entropy flux from the system to the reservoir as $ J_S(\rho) = -\beta \partial_t \rho (H_S)$, one sees that (\ref{entprod}) gives nothing else than the entropy production $\sigma(\rho) = \partial_t S(\rho)+ J_S(\rho) =   \partial_t \rho( \Delta S)$, analogously to what already remarked in~\cite{Ottinger2010_TLS_DHO}. Since $\sigma(\rho)$ is strictly positive unless $\Delta S=c 1$ for a $c$-number (due to our ergodic hypothesis), this implies both the uniqueness and the thermodynamic stability of the steady state $\rho_+=\rho_\beta$. 

In the belief that the ergodicity assumption should imply that the steady state $\rho_+$ is unique, when it exists, also for the case of $N$ heat baths at inverse temperatures $\beta_j$, $j=1\ldots,N$, we now define the Onsager matrix~\cite{DegrMaz} in the linear regime, where the entropy fluxes are linearly proportional to the thermodynamic forces $X_j=\beta_j-\beta$ (for a reference inverse temperature $\beta$). We prove that the Onsager matrix is positive and give a Green-Kubo formula~\cite{Kubo1991,Jakia_2006kl} for its coefficients (thereby obtaining the Onsager symmetry relations). These ideas and techniques combine the analyses in~\cite{Grabert1982,PhysRevA.90.042110} for a time-driven setting, and in~\cite{SpohnLeb} for the Davies case.

To this end, we consider the MDS $\partial_t\rho = D(\rho) = \rho\left( i[H_S,\cdot]\right) +\sum_{j=1}^N D^{\rm d}_{j,\beta_j}(\rho)$ where each dissipative contribution $D^{\rm d}_{j,\beta_j}(\rho) =  \rho\left([[\cdot,\Delta S_j]]_{\beta_j,\rho}^{j} \right)$ is given in (\ref{MDS}) with scattering operators $Q^\lambda_{j,\beta_j}$, and relative entropy operator $\Delta S_j=\Delta S(\beta_j) = \ln\rho - \ln\rho_{\beta_j}$. Note that $\rho_\beta$ is the only steady state at equilibrium $\beta_1=\ldots=\beta_N=\beta$ (as can be seen from the above entropy production argument) and that, under our ergodic assumption, $\rho_{\beta_j}$ is the unique steady state of $D^{\rm d}_{j,\beta_j}(\cdot)$. 

Using $K_\rho A = \int_0^1 \rho^\lambda A \rho^{1-\lambda}\, d\lambda$ (which admits an inverse) we compute the linearization $D^{\rm d}_{j,\beta}(\rho) = \bar D^{\rm d}_{j,\beta}  \rho +o(\rho - \rho_{\beta})$ at the equilibrium~\cite{foot2} to be
\begin{equation}\label{linear}
\bar D^{\rm d}_{j,\beta} \rho  = - \rho_{\beta}([[\cdot, K^{-1}_{\rho_{\beta}}\rho]]^{j}_{\beta,\rho_{\beta}} ).
\end{equation}
We see that $\bar D^{\rm d}_{j,\beta}$ is purely dissipative, e.g. $ \bar D^{\rm d}_{j,\beta} \rho= \omega \rho$ only has solutions with $\omega \leq 0$, and that $\rho_{\beta}$ is the unique steady state of $\bar D^{\rm d}_{j,\beta}$. Indeed by tracing the eigenvalue equation against $K^{-1}_{\rho_\beta}\rho$, the right hand side reads $ \omega\, {\rm tr} (\rho K^{-1}_{\rho_\beta} \rho)$, and the trace is positive, as follows from a calculation in the $H_S$ basis. The left hand side is the negative of
\begin{equation}\label{entrprlin}
\rho_{\beta}([[K^{-1}_{\rho_{\beta}}\rho, K^{-1}_{\rho_{\beta}}\rho]]^{j}_{\beta,\rho_{\beta}} )
\end{equation}
and is thus strictly negative unless $\rho = \rho_\beta$. The argument follows. Note that $\bar D^{\rm d}_{j,\beta}$ need not be of Lindblad type.

The entropy production at the steady state is $\sigma(\rho_+) = \sum_{j=1}^N \sigma_j(\rho_+) =  \sum_{j=1}^N X_j J_j $ where the mean fluxes $J_j=\rho_+( [[H_S, \Delta S_k(\rho_+)]]^{j}_{\beta_j,\rho_+})=\sum_{k=1}^N X_k L_{jk}(\beta) +o(X)$ define the Onsager coefficients $L_{jk}(\beta)$. 
From the above linearization we deduce that the entropy production $\sigma(\rho_+) $ next to equilibrium is a bilinear form in $\rho_+ - \rho_\beta$, whose terms $\sigma_j(\rho_+)$ are precisely given by (\ref{entrprlin}) with $\beta=\beta_j$. In particular $\rho_\beta$ is a local minimum for $\sigma(\rho_+)$, which implies that the Onsager matrix is positive definite (over the reals). 
To prove the Onsager relations $L_{kj}(\beta) = L_{jk}(\beta)$ we need a notion of detailed balance next to equilibrium. To achieve this, we note that we can safely trade the mean fluxes $J_j$ with their linearization $\bar J_j$ made at $\beta_j$ (according to (\ref{linear})), and thus limit ourselves to analyze the detailed balance properties of $\bar D^{\rm d}=\sum_{j=1}^N \bar D^{\rm d}_{j,\beta_j}$ at equilibrium. Defining $\bar D = -i[H_S,\cdot]+\bar D^{\rm d}$ and $\bar D^{*}$ on $\mathcal B(\mathcal H)$ by duality, we note that at equilibrium 
\begin{equation}\label{db}
\bar D^{*} = K^{-1}_{\rho_\beta} \bar D K_{\rho_\beta}.
\end{equation}
We remark that this detailed balance property is contained in the modular structure of the MDS, and does not come a priori from the KMS property~\cite{Kossakowski1977} of some bath correlation function~\cite{Davies1974}. Of course, these two notions can be linked through our specific choice of the coupling operator $Q^\lambda$, as we shall see, but the conceptual separation is evident. We are now in position to go through Lemma 1 of~\cite{SpohnLeb}, with the only difference that we now get $\bar D^{\rm d}_{j,\beta}(H_S\rho_\beta) = \bar D^{\rm d}_{j,\beta}(K_{\rho_\beta}H_S) = K_{\rho_\beta}\bar D^{{\rm d},*}_{j,\beta}(H_S)$ at equilibrium. By defining the flux operators $\mathbb J_j = \bar D^{{\rm d},*}_{j,\beta}(H_S)$, and using the fact that $\bar D$ is dissipative (as can be similarly proven for $\bar D^*$ as well) the Green-Kubo formula is obtained
\begin{equation}\label{onsrel}
L_{jk}(\beta) = \int_0^\infty \!\!\!\!\langle  \mathbb J_j (t) \: ; \:   \mathbb J_k \rangle_\beta dt. 
\end{equation}
The r.h.s. is written with the help of the Kubo scalar product $\langle  A \: ; \:  B \rangle_\beta = {\rm tr} (A^\dagger K_{\rho_\beta} B)$ and the Heisenberg evolution $A(t) = e^{\bar D^{*}  t} A$. The detailed balance property $(\ref{db})$ implies that $  \mathbb J_j (t) = K^{-1}_{\rho_\beta} e^{\bar D  t} (K_{\rho_\beta}  \mathbb J_j)$, from which it follows that $\langle  \mathbb J_j (t) \: ; \:  \mathbb J_k \rangle_\beta = \langle  \mathbb J_k (t) \: ; \:  \mathbb J_j \rangle_\beta$, providing the Onsager symmetry relations. Note that the currents $  \mathbb J_j$ need not commute with $\rho_\beta$, so our formula is in general different from the one provided in~\cite{SpohnLeb}, and is exactly the well known Kubo formula~\cite{Kubo1991}.

The analysis above puts the MDS on solid thermodynamical grounds. We shall now show that the set of interesting MDS is far from empty. The first notable example is the celebrated Davies generator~\cite{Davies1974}. To see that it belongs to our MDS class, take eigenoperators $A_{\nu}$ defined by $[A_\nu,H_S] = \nu A_\nu$, a bath spectral function $\hat h(\nu)>0$ with the KMS property $\hat h(\nu) = e^{\beta \nu} \hat h(-\nu)$~\cite{Davies1974}, and define $Q_{\nu}^\lambda = e^{-\lambda\beta \nu/2} \sqrt{\hat h(\nu)} A_\nu$. Then (\ref{MDS}) is generalized to the sum over all $\nu$ of the corresponding brackets $[[\cdot,\cdot]]^\nu_{\beta,\rho}$, and is guaranteed to be self-adjoint by the natural generalization $Q_{\nu}^{\lambda \dagger} = Q^{1-\lambda}_{-\nu}$ to many coupling operators $Q^\lambda_{\nu}$. The resulting master equation reads
\begin{equation}\label{davies}
\partial_t \rho = -i[H_S,\rho] +\sum_\nu \hat h(\nu)
 \left(-{1\over 2}\{ A_{\nu}^\dagger A_{\nu},\rho \} + A_{\nu}\rho A_{\nu}^\dagger\right).
\end{equation}
This can be seen from $
\int_0^1 e^{-\lambda\beta\nu} \rho^\lambda [A_{\nu},F]\rho^{1-\lambda} \, d\lambda =A_{\nu}\rho -e^{-\beta\nu}\rho A_{\nu} $, 
as can be computed in the $\rho$ basis, and then using the KMS property of $\hat h$. The r.h.s. of Eq. (\ref{davies}) gives exactly the Davies generator~\cite{Davies1974}, describing the weak coupling limit of a subsystem $S$ coupled to a thermal bath $B$ at inverse temperature $\beta$ through a hamiltonian $H=H_S\otimes 1 + R\otimes \Phi + 1\otimes H_B$, where $\hat h(\nu)$ is Fourier related to $ h(t)  =\langle e^{i H_B t}\Phi e^{-iH_B t}\Phi\rangle_{\beta}$ and $R = \sum_\nu A_\nu$. In passing, we remark that the Davies generator arising from a more general coupling hamiltonian $\sum_\alpha R_\alpha\otimes \Phi_\alpha$ can just as well be cast into MDS form, through a trivial extension of the above with eigenoperators $A_{\alpha}$ relative to frequencies $\nu_\alpha$. We note that the KMS property of the bath correlation functions is here embedded as a necessary condition to have the MDS well defined: This is the natural generalization $Q_{\nu}^{\lambda \dagger} = Q^{1-\lambda}_{-\nu}$ of the self-adjointness condition $Q^{\lambda\dagger} = Q^{1-\lambda}$ to many coupling operators $Q^\lambda_{\nu}$. That the linearization of the Davies generator according to (\ref{linear}) is again the Davies generator was already proven in~\cite{PhysRevA.90.042110}. Since in this case $[\mathbb J_j,H_S]=0$, our expressions for the Onsager coefficients agrees with those computed in~\cite{SpohnLeb}. 

The above unravels the thermodynamic and modular structure of the Davies generator, and automatically gives different proofs and more symmetric expressions for the entropy production, the fluxes and the Onsager relations. We make the following remarks. The boundedness of the relative entropy operator is more relevant to a thermodynamic setting than complete positivity. The former is guaranteed by the modular structure in (\ref{MDS}), while the latter by the Lindblad structure~\cite{Lindblad1976}. Second, from the modular perspective, the linearity of the Davies generator, together with its complete positivity, appears to be a fortunate, accidental coincidence, which is due to the peculiar commutation relations of the eigenoperators $A_{\nu}$ appearing in the coupling operator $Q^\lambda_\nu$. The eigenoperators exist however only when $H_S$ has a purely discrete spectrum, which severely limits applications to systems without internal structure, namely quantum dots and harmonic oscillators. Moreover even for those, other alternative MDS could be of interest.

These remarks led us to a previous work by one of us~\cite{Taj2010}, where the eigenoperators $A_\nu$ were traded with different scattering operators  
\begin{equation}
\tilde A_{\nu} = \int_{-\infty}^{+\infty} e^{i\nu t} \sqrt{\delta(t,T)}\, e^{iH_S t} R e^{iH_S t} \, dt,
\end{equation} 
where $\nu\in\mathbb R$, $\delta(t,T)$ is the normalized Gaussian at $t$, whose standard deviation $T$ physically represents a collision time (which can depend on a coupling constant, in a weak coupling limit spirit). Contrary to the eigenoperators, the $\tilde A_\nu$'s are always well defined and bounded (as long as $0<T< \infty$), independently of the spectral properties of $H_S$. Proceeding as before, we now take our coupling operators to be $\tilde Q_{\nu}^\lambda = e^{-\lambda\beta \nu/2} \sqrt{\hat h(\nu)} \tilde A_\nu$, and with those, we define the dissipative brackets $\tilde{[[}\cdot,\cdot\tilde{]]}_{\beta,\rho}^\nu$ like in~(\ref{disbra}). The resulting MDS master equation 
\begin{equation}\label{neweq}
\partial_t\rho(\cdot) = \rho\left( i[H_S,\cdot]\right) + \rho\left(\int_{\mathbb R} \tilde{[[}\cdot,\Delta S\tilde{]]}^\nu_{\beta,\rho} d\nu\right)
\end{equation}
is still well defined ($\tilde A_\nu^\dagger = \tilde A_{-\nu}$ implies $\tilde Q_{\nu}^{\lambda \dagger} = \tilde Q^{1-\lambda}_{-\nu}$) and does not suffer from any spectral restriction on the system Hamiltonian. However, the above MDS master equation will in general be genuinely nonlinear. It can be seen with analogous calculations as in~\cite{Taj2010} that for $H_S$ with discrete spectrum it boils down to the Davies generator when $T\rightarrow \infty$. As before, the analysis easily accommodates for more general coupling hamiltonians $\sum_\alpha R_\alpha\otimes \Phi_\alpha$. It follows that (\ref{neweq}) is fully justified on the statistical mechanical basis of the weak coupling limit, and  provides a genuine and thermodynamically robust MDS extension of the Davies master equation, able to describe even unbounded systems, possibly in the thermodynamic limit.

We also remark that when the spectral function $\hat h(\nu)$ is singular at $\nu=0$ (white noise), then the only coupling operator involved in (\ref{neweq}) is (possibly some $n$-th order derivative of) $\tilde Q^\lambda_{\nu}$ at $\nu=0$, which will not depend upon $\lambda$. In this case, (\ref{neweq}) gives back the nonlinear master equation in~\cite{Ottinger2010_TLS_DHO,Ottinger2011_GEO}.  This raises the observation that (\ref{neweq}) should be regarded as a (nonlinear) quantum counterpart of the classical linear Boltzmann equation, taking the singular limit of the scattering rates of which~\cite{risken} gives the (nonlinear) quantum counterpart~\cite{Ottinger2010_TLS_DHO} of the classical Fokker-Planck equation. 

We finally want to address the problem of unicity of the MDS fulfilling the properties (i-vi) stated in the introduction. Although we have already proven the existence of two MDS master equations for bounded systems ((\ref{davies}) and (\ref{neweq})), many more possibilities remain. As an additional example, one could consider a single coupling operator $Q^\lambda = \sum_\nu e^{-\lambda\beta \nu/2} \sqrt{\hat h(\nu)} A_\nu$, and the resulting nonlinear evolution will still match the corresponding coupled hamiltonian evolution in the weak coupling limit. To see that, it is sufficient to take the time average map of the MDS master equation~\cite{Davies1974} and realize that the result is just (\ref{davies}) (a solid argument as to why this should imply agreement in the weak coupling limit can be found in~\cite{Davies1976}). However, while these MDS all agree in a limit regime, they will in general give (possibly qualitative) different predictions at small but finite values of the coupling to the bath, which is where master equations are used in applications.

To summarize, we have proposed the class of MDS master equations, describing the Markovian, nonlinear dissipative dynamics of a quantum system interacting with one or more particle reservoirs at equilibrium. The MDS is structurally and thermodynamically robust, as opposed to a general QDS. From this perspective, we find that the Davies master equation generates the only MDS which, accidentally, also happens to be a QDS. This is just due to the peculiar character of the corresponding coupling operators. We have used the MDS structure of the Davies generator to extend the latter to a more general, nonlinear MDS master equation, which is free form the severe restrictions of the former and can even describe unbounded systems, possibly in the thermodynamical limit. Other possible MDSs agree with the Davies MDS in the weak coupling limit, while differing from it at finite values of the coupling to the bath. The origin of the nonlinearity of the MDS lies in the intrinsic entropic dependence of the scattering event, and is very different in nature from any mean field approach, where nonlinearities are normally expected to arise. The MDS master equation relative to state-independent coupling operators should be regarded as the quantum counterpart of the classical linear Boltzmann equation.

We would like to thank M. Osmanov and J. Flakowski for their invaluable support and stimulating discussions.



\bibliography{MDS_biblio}

\begin{thebibliography}{28}
\expandafter\ifx\csname natexlab\endcsname\relax\def\natexlab#1{#1}\fi
\expandafter\ifx\csname bibnamefont\endcsname\relax
  \def\bibnamefont#1{#1}\fi
\expandafter\ifx\csname bibfnamefont\endcsname\relax
  \def\bibfnamefont#1{#1}\fi
\expandafter\ifx\csname citenamefont\endcsname\relax
  \def\citenamefont#1{#1}\fi
\expandafter\ifx\csname url\endcsname\relax
  \def\url#1{\texttt{#1}}\fi
\expandafter\ifx\csname urlprefix\endcsname\relax\def\urlprefix{URL }\fi
\providecommand{\bibinfo}[2]{#2}
\providecommand{\eprint}[2][]{\url{#2}}

\bibitem[{\citenamefont{\"{O}ttinger}(2011)}]{Ottinger2011_GEO}
\bibinfo{author}{\bibfnamefont{H.~C.} \bibnamefont{\"{O}ttinger}},
  \bibinfo{journal}{Europhys.~Lett.} \textbf{\bibinfo{volume}{94}},
  \bibinfo{pages}{10006} (\bibinfo{year}{2011}).

\bibitem[{\citenamefont{\"{O}ttinger}(2010)}]{Ottinger2010_TLS_DHO}
\bibinfo{author}{\bibfnamefont{H.~C.} \bibnamefont{\"{O}ttinger}},
  \bibinfo{journal}{Phys. Rev. A} \textbf{\bibinfo{volume}{82}},
  \bibinfo{pages}{052119} (\bibinfo{year}{2010}).

\bibitem[{\citenamefont{Davies}(1974)}]{Davies1974}
\bibinfo{author}{\bibfnamefont{E.}~\bibnamefont{Davies}},
  \bibinfo{journal}{Comm. Math. Phys.} \textbf{\bibinfo{volume}{39}},
  \bibinfo{pages}{91} (\bibinfo{year}{1974}).

\bibitem[{\citenamefont{Gorini et~al.}(1978)\citenamefont{Gorini, Frigerio,
  Verri, Kossakowski, and Sudarshan}}]{Gorini1978149}
\bibinfo{author}{\bibfnamefont{V.}~\bibnamefont{Gorini}},
  \bibinfo{author}{\bibfnamefont{A.}~\bibnamefont{Frigerio}},
  \bibinfo{author}{\bibfnamefont{M.}~\bibnamefont{Verri}},
  \bibinfo{author}{\bibfnamefont{A.}~\bibnamefont{Kossakowski}},
  \bibnamefont{and}
  \bibinfo{author}{\bibfnamefont{E.}~\bibnamefont{Sudarshan}},
  \bibinfo{journal}{Rep.~Math.~Phys.} \textbf{\bibinfo{volume}{13}},
  \bibinfo{pages}{149 } (\bibinfo{year}{1978}).

\bibitem[{\citenamefont{Alicki and Lendi}(2007)}]{Alicki2007}
\bibinfo{author}{\bibfnamefont{R.}~\bibnamefont{Alicki}} \bibnamefont{and}
  \bibinfo{author}{\bibfnamefont{K.}~\bibnamefont{Lendi}}, in
  \emph{\bibinfo{booktitle}{Lect.~Notes~Phys.}}
  (\bibinfo{publisher}{Springer-Verlag}, \bibinfo{year}{2007}), vol.
  \bibinfo{volume}{717}.

\bibitem[{\citenamefont{Breuer and Petruccione}(2002)}]{Breuer2002}
\bibinfo{author}{\bibfnamefont{H.-P.} \bibnamefont{Breuer}} \bibnamefont{and}
  \bibinfo{author}{\bibfnamefont{F.}~\bibnamefont{Petruccione}},
  \emph{\bibinfo{title}{The {T}heory of {O}pen {Q}uantum {S}ystem}}
  (\bibinfo{publisher}{Oxford University Press}, \bibinfo{year}{2002}).

\bibitem[{\citenamefont{Davies}(1976)}]{Davies1976}
\bibinfo{author}{\bibfnamefont{E.}~\bibnamefont{Davies}},
  \bibinfo{journal}{Ann.~Math.} \textbf{\bibinfo{volume}{219}},
  \bibinfo{pages}{147} (\bibinfo{year}{1976}).

\bibitem[{\citenamefont{Davies}(1975)}]{Davies1975}
\bibinfo{author}{\bibfnamefont{E.~B.} \bibnamefont{Davies}},
  \emph{\bibinfo{title}{Quantum Theory of Open Quantum Systems}}
  (\bibinfo{publisher}{Academic Press, London}, \bibinfo{year}{1975}).

\bibitem[{\citenamefont{Spohn and Lebowitz}(1978)}]{SpohnLeb}
\bibinfo{author}{\bibfnamefont{H.}~\bibnamefont{Spohn}} \bibnamefont{and}
  \bibinfo{author}{\bibfnamefont{J.~L.} \bibnamefont{Lebowitz}},
  \bibinfo{journal}{Adv.~Chem.~Phys.} \textbf{\bibinfo{volume}{38}}
  (\bibinfo{year}{1978}).

\bibitem[{\citenamefont{Spohn}(1978)}]{Spohn1978}
\bibinfo{author}{\bibfnamefont{H.}~\bibnamefont{Spohn}}, \bibinfo{journal}{J.
  Math. Phys.} \textbf{\bibinfo{volume}{19}}, \bibinfo{pages}{1227}
  (\bibinfo{year}{1978}).

\bibitem[{\citenamefont{Alicki}(1977)}]{Alicki_1977eu}
\bibinfo{author}{\bibfnamefont{R.}~\bibnamefont{Alicki}},
  \bibinfo{journal}{Int.~J.~Theor.~Phys.} \textbf{\bibinfo{volume}{16}},
  \bibinfo{pages}{351} (\bibinfo{year}{1977}).

\bibitem[{\citenamefont{Lindblad}(1976)}]{Lindblad1976}
\bibinfo{author}{\bibfnamefont{G.}~\bibnamefont{Lindblad}},
  \bibinfo{journal}{Comm. Math. Phys.} \textbf{\bibinfo{volume}{48}},
  \bibinfo{pages}{119} (\bibinfo{year}{1976}).

\bibitem[{\citenamefont{Taj}(2010)}]{Taj2010}
\bibinfo{author}{\bibfnamefont{D.}~\bibnamefont{Taj}},
  \bibinfo{journal}{Ann.~Henri~Poincar\'e} \textbf{\bibinfo{volume}{11}},
  \bibinfo{pages}{1303} (\bibinfo{year}{2010}).

\bibitem[{\citenamefont{Gorini et~al.}(1976)\citenamefont{Gorini, Kossakowski,
  and Sudarshan}}]{gorini76}
\bibinfo{author}{\bibfnamefont{V.}~\bibnamefont{Gorini}},
  \bibinfo{author}{\bibfnamefont{A.}~\bibnamefont{Kossakowski}},
  \bibnamefont{and} \bibinfo{author}{\bibfnamefont{E.~C.~G.}
  \bibnamefont{Sudarshan}}, \bibinfo{journal}{J.~Math.~Phys.}
  \textbf{\bibinfo{volume}{17}}, \bibinfo{pages}{821} (\bibinfo{year}{1976}).

\bibitem[{\citenamefont{De~Groot and Mazur}(1962)}]{DegrMaz}
\bibinfo{author}{\bibfnamefont{S.~R.} \bibnamefont{De~Groot}} \bibnamefont{and}
  \bibinfo{author}{\bibfnamefont{P.}~\bibnamefont{Mazur}},
  \emph{\bibinfo{title}{Non-Equilibrium Thermodynamics}}
  (\bibinfo{publisher}{North-Holland, Amsterdam}, \bibinfo{year}{1962}).

\bibitem[{\citenamefont{Kubo et~al.}(1991)\citenamefont{Kubo, Toda, and
  Hashitsume}}]{Kubo1991}
\bibinfo{author}{\bibfnamefont{R.}~\bibnamefont{Kubo}},
  \bibinfo{author}{\bibfnamefont{M.}~\bibnamefont{Toda}}, \bibnamefont{and}
  \bibinfo{author}{\bibfnamefont{N.}~\bibnamefont{Hashitsume}},
  \emph{\bibinfo{title}{Nonequilibrium {S}tatistical {M}echanics}},
  vol.~\bibinfo{volume}{II} (\bibinfo{publisher}{Springer-Verlag},
  \bibinfo{year}{1991}).

\bibitem[{\citenamefont{Jaksic et~al.}(2006)\citenamefont{Jaksic, Ogata, and
  Pillet}}]{Jakia_2006kl}
\bibinfo{author}{\bibfnamefont{V.}~\bibnamefont{Jaksic}},
  \bibinfo{author}{\bibfnamefont{Y.}~\bibnamefont{Ogata}}, \bibnamefont{and}
  \bibinfo{author}{\bibfnamefont{C.}~\bibnamefont{Pillet}},
  \bibinfo{journal}{Comm.~Math.~Phys.} \textbf{\bibinfo{volume}{265}},
  \bibinfo{pages}{721} (\bibinfo{year}{2006}).

\bibitem[{\citenamefont{Taj and Rossi}(2008)}]{Taj2008}
\bibinfo{author}{\bibfnamefont{D.}~\bibnamefont{Taj}} \bibnamefont{and}
  \bibinfo{author}{\bibfnamefont{F.}~\bibnamefont{Rossi}},
  \bibinfo{journal}{Phys. Rev. A} \textbf{\bibinfo{volume}{78}},
  \bibinfo{pages}{052113} (\bibinfo{year}{2008}).

\bibitem[{\citenamefont{Bratteli and Robinson}(1979)}]{bratteli}
\bibinfo{author}{\bibfnamefont{O.}~\bibnamefont{Bratteli}} \bibnamefont{and}
  \bibinfo{author}{\bibfnamefont{D.~W.} \bibnamefont{Robinson}},
  \emph{\bibinfo{title}{Operator Algebras and Quantum Statistical Mechanics}},
  vol.~\bibinfo{volume}{I} (\bibinfo{publisher}{Springer-Verlag, Berlin},
  \bibinfo{year}{1979}).

\bibitem[{\citenamefont{Jaksic et~al.}(2010)\citenamefont{Jaksic, Ogata,
  Pautrat, and Pillet}}]{Pillet2010}
\bibinfo{author}{\bibfnamefont{V.}~\bibnamefont{Jaksic}},
  \bibinfo{author}{\bibfnamefont{Y.}~\bibnamefont{Ogata}},
  \bibinfo{author}{\bibfnamefont{Y.}~\bibnamefont{Pautrat}}, \bibnamefont{and}
  \bibinfo{author}{\bibfnamefont{C.-A.} \bibnamefont{Pillet}}, in
  \emph{\bibinfo{booktitle}{Quantum Theory from Small to Large Scales}}
  (\bibinfo{year}{2010}), vol.~\bibinfo{volume}{95}, pp.
  \bibinfo{pages}{213--410}.

\bibitem[{\citenamefont{Jaksic et~al.}(2013)\citenamefont{Jaksic, Pillet, and
  Westrich}}]{Jaksic2013_QDS}
\bibinfo{author}{\bibfnamefont{V.}~\bibnamefont{Jaksic}},
  \bibinfo{author}{\bibfnamefont{C.-A.} \bibnamefont{Pillet}},
  \bibnamefont{and} \bibinfo{author}{\bibfnamefont{M.}~\bibnamefont{Westrich}},
  \bibinfo{journal}{J.~Stat.~Phys.} pp. \bibinfo{pages}{1--35}
  (\bibinfo{year}{2013}).

\bibitem[{\citenamefont{Alicki and Majewski}(1990)}]{alicki1990}
\bibinfo{author}{\bibfnamefont{R.}~\bibnamefont{Alicki}} \bibnamefont{and}
  \bibinfo{author}{\bibfnamefont{W.~A.} \bibnamefont{Majewski}},
  \bibinfo{journal}{Phys.~Lett.~A} \textbf{\bibinfo{volume}{148}},
  \bibinfo{pages}{69} (\bibinfo{year}{1990}).

\bibitem[{\citenamefont{Shaji and Sudarshan}(2005)}]{Shaji200548}
\bibinfo{author}{\bibfnamefont{A.}~\bibnamefont{Shaji}} \bibnamefont{and}
  \bibinfo{author}{\bibfnamefont{E.}~\bibnamefont{Sudarshan}},
  \bibinfo{journal}{Phys.~Lett.~A} \textbf{\bibinfo{volume}{341}},
  \bibinfo{pages}{48 } (\bibinfo{year}{2005}), ISSN \bibinfo{issn}{0375-9601}.

\bibitem[{\citenamefont{Grabert}(1982)}]{Grabert1982}
\bibinfo{author}{\bibfnamefont{H.}~\bibnamefont{Grabert}},
  \bibinfo{journal}{Z.~Physik~B} \textbf{\bibinfo{volume}{49}},
  \bibinfo{pages}{161} (\bibinfo{year}{1982}).

\bibitem[{\citenamefont{Flakowski et~al.}(2014)\citenamefont{Flakowski,
  Osmanov, Taj, and \"Ottinger}}]{PhysRevA.90.042110}
\bibinfo{author}{\bibfnamefont{J.}~\bibnamefont{Flakowski}},
  \bibinfo{author}{\bibfnamefont{M.}~\bibnamefont{Osmanov}},
  \bibinfo{author}{\bibfnamefont{D.}~\bibnamefont{Taj}}, \bibnamefont{and}
  \bibinfo{author}{\bibfnamefont{H.~C.} \bibnamefont{\"Ottinger}},
  \bibinfo{journal}{Phys.~Rev.~A} \textbf{\bibinfo{volume}{90}},
  \bibinfo{pages}{042110} (\bibinfo{year}{2014}).

\bibitem[{foo()}]{foot2}
\bibinfo{note}{This follows from $\ln\rho - \ln\rho_{\beta} =
  K^{-1}_{\rho_{\beta}}\delta\rho + o(\delta\rho)$ for $\delta\rho =
  \rho-\rho_{\beta}$.}

\bibitem[{\citenamefont{Kossakowski et~al.}(1977)\citenamefont{Kossakowski,
  Frigerio, Gorini, and Verri}}]{Kossakowski1977}
\bibinfo{author}{\bibfnamefont{A.}~\bibnamefont{Kossakowski}},
  \bibinfo{author}{\bibfnamefont{A.}~\bibnamefont{Frigerio}},
  \bibinfo{author}{\bibfnamefont{V.}~\bibnamefont{Gorini}}, \bibnamefont{and}
  \bibinfo{author}{\bibfnamefont{M.}~\bibnamefont{Verri}},
  \bibinfo{journal}{Comm. Math. Phys.} \textbf{\bibinfo{volume}{57}},
  \bibinfo{pages}{97} (\bibinfo{year}{1977}).

\bibitem[{\citenamefont{Risken}(1996)}]{risken}
\bibinfo{author}{\bibfnamefont{H.}~\bibnamefont{Risken}},
  \emph{\bibinfo{title}{The Fokker-Planck Equation}}, vol.~\bibinfo{volume}{18}
  (\bibinfo{publisher}{Springer Series in Synergetics}, \bibinfo{year}{1996}).

\end{thebibliography}



\end{document}